\documentclass{aa}
\usepackage{graphicx}
\usepackage{txfonts}
\usepackage{natbib}
\usepackage{lscape}
\usepackage{xcolor}     

\bibpunct{(}{)}{;}{a}{}{,}
\usepackage{float}
\usepackage{appendix}
\usepackage{xcolor}
\usepackage{amsmath}
\usepackage{hyperref}

\begin{document} 

   \title{Discovery and characterization of ZTF J0112+5827: An 80.9-minute polar with strong cyclotron features} 

   \subtitle{}

   \author{Jiamao Lin\inst{1,2}\fnmsep\thanks{ Corresponding author.}, 
    Liangliang Ren\inst{3}, Chengyuan Li\inst{1,2}, Elias-Rosa Nancy\inst{4,5}, Tianqi Cang\inst{6}, Hongwei Ge\inst{7,8}, Pak-Hin Thomas Tam\inst{1,2}, Wenjun Huang\inst{1,2}, Yilong Li\inst{1,2}, Xiaofeng Wang\inst{9,10}, Yang Huang\inst{11,12} and Bo Ma\inst{1,2}}
    \institute{School of Physics and Astronomy, Sun Yat-sen University, Zhuhai 519082, China \\
    \email{linjm66@mail2.sysu.edu.cn} \and
    CSST Science Center for the Guangdong-Hongkong-Macau Greater Bay Area, Sun Yat-sen University, Zhuhai 519082, China \and
    School of Electrical and Electronic Engineering, Anhui Science and Technology University, Bengbu, Anhui 233030, China \and
    INAF-Osservatorio Astronomico di Padova, Vicolo dellâ’Osservatorio 5, 35122 Padova, Italy 
    \and Institute of Space Sciences (ICE), Consejo Superior de Investigaciones CientÃíficas (CSIC), Barcelona, Spain \\
    \email{nancy.elias@inaf.it}\and
    School of Physics and Astronomy, Beijing Normal University, Beijing 100875, China \and
    Yunnan Observatories, Chinese Academy of Sciences, Kunming 650216, China 
    \and
    International Center of Supernovae, Yunnan Key Laboratory, Kunming 650216, China \and
    Physics Department and Tsinghua Center for Astrophysics, Tsinghua University, Beijing 100084, China \and
    Beijing Planetarium, Beijing Academy of Science and Technology, Beijing 100044, China \and
    University of Chinese Academy of Sciences, Beijing 100049, China\and 
    National Astronomical Observatories, Chinese Academy of Sciences, Beijing 100012, China}

   \date{Submitted: \today }

 
  \abstract
   {A new X-ray Cataclysmic variable (CV) candidate exhibits distinct light-curve characteristics in the ZTF's $g$, $r$, and $i$ bands. The paper includes the optical identification and multiwavelength analysis of this CV candidate.}
   {This work aims to determine if a previously identified CV candidate, ZTF J0112+5827, is a polar system by examining its X-ray and cyclotron radiation characteristics.
   }
    {We characterized the X-ray emission of ZTF J0112+5827 using the ROSAT observations. The $gri$-band optical light curves were obtained from the \textit{Zwicky} Transient Facility. After two nights of time-domain spectroscopic observations with the Palomar 200-inch telescope, we mapped the accretion structures using Doppler tomography.}
   {ZTF J0112+5827 exhibits an orbital period of 80.9 minutes, determined from the ZTF light curves, and an average X-ray flux of $(68.4 \pm 15.7) \times 10^{-14}$ erg s$^{-1}$ cm$^{-2}$ in the 0.1–2.4 keV range. It shows an ellipsoidal-like variability curve in the $g$ band, with two prominent humps around phases of $\sim$0.0 and $\sim$0.7 in the $i$ and $r$ bands. In the spectra corresponding to these phases, a redward-increasing power-law continuum appears, which is accompanied by prominent features of cyclotron emission humps. Emission lines of He II and Balmer series were observed. The magnetic field strength of ZTF J0112+5827 was determined from the cyclotron harmonics. Its tomography map revealed the presence of accretion streams, 
   but there was no evidence of an accretion disk structure. The line-of-sight velocity of the Balmer emission was measured at about 500 km s$^{-1}$, the majority of which was contributed by accretion streams and accretion spots. Our result confirms that ZTF J0112+5827 is a polar system. It contains a magnetic white dwarf with a magnetic field strength of $ 38.7_{-1.1}^{+1.3} MG$.}
   
   {}
   \keywords{stars: cataclysmic variables - gravitational waves - magnetic field: white dwarfs}
   \titlerunning{ZTF J0112+5827}
   \authorrunning{Lin et al.}

   \maketitle
%

\section{Introduction}

Cataclysmic variable stars (CVs) are binary systems where a white dwarf (WD) accretes material from a companion star that exceeds its Roche lobe \citep[e.g.,][]{warner95,hellierbook,knigge2011}. These systems are formed through a common-envelope event, where the more massive star evolves off the main sequence (MS) and expands, engulfing its companion. During the common envelope phase, the binary system dissipates a significant amount of angular momentum due to friction, resulting in a rapid contraction of the orbital semi-major axis. Studying CVs is fundamentally significant as it serves as a testing ground for theories relating to accretion, which can then be applied across various realms of astrophysics, including star formation, galaxy formation, and active galactic nuclei. Recently, AM Canum Venaticorum (AM CVn) stars, which are important gravitational-wave (GW) sources for both the Laser Interferometer Space Antenna \citep[LISA,][]{2017amaro} and Tianqin \citep{2020PhRvD.102f3021H}, were found to be potentially more abundant when formed through the evolved CV channel \citep{2023sarkar}.

Among CVs, polars are a unique subclass characterized by a WD with a strong magnetic field ($>$10 MG) that synchronizes its rotation with the binary orbital period. The accretion flow from the donor star initially follows a ballistic trajectory until it comes under the influence of the magnetic pressure from the WD. This magnetic interaction diverts the accretion flow, channeling it along the magnetic field lines toward a cyclotron-emitting accretion region near the WD's magnetic poles. As this gas accelerates toward the WD at supersonic speeds, it encounters a shock front near the stellar surface, which generates extremely high temperatures. Upon crossing this shock front, the gas decelerates and cools as it settles onto the WD's surface, emitting bremsstrahlung X-rays and optical cyclotron radiation as it radiates away the excess energy. This magnetic channeling effectively inhibits the formation of an accretion disk \citep[e.g.,][]{1990cropper, mukai2017}.

Studying polars helps explain how magnetic fields alter the angular momentum evolution of binary stars. The standard model of nonmagnetic CVs' evolution posits that when the donor star becomes fully convective at an orbital period of approximately three hours, magnetic braking (MB) abruptly stops. The MB leads to a halt in accretion, resulting in a notable lack of active CVs within the period range of two to three hours, which is known as the CV period gap  \citep{1971faulkner,1981verbunt,rappaport1983}. However, the period distribution of polars does not exhibit this gap \citep{2017hameury_dn_ip,schreiber2021}. In nonmagnetic CVs with minimum periods below approximately 80 minutes, the donor star becomes degenerate and expands as it loses mass, resulting in a period bounce \citep[]{1999kolb,2009gansicke}. Polars exhibit shorter minimum periods compared to nonmagnetic CVs. \cite{2024schreiber} suggest that the WD's magnetic field significantly diminishes the wind loss from the secondary star, leading to a deceleration in angular-momentum loss. Nevertheless, the current sample size of observed magnetic CVs remains limited, making it difficult to obtain a comprehensive understanding of the evolution of systems such as polars.

Polars may also be potential candidates for WDs exceeding the Chandrasekhar limit. Recent observations have detected a number of Type Ia supernovae with exceptionally high luminosity, indicating progenitor WDs masses significantly exceeding the Chandrasekhar mass limit. In the presence of a strong magnetic field, stellar equilibrium configurations can be supported by Lorentz forces, enabling the support of greater mass.  This theoretical framework is considered a possible explanation for the existence of ultra-massive WDs. \citep[]{2006andrewhowell,2013das,2017bera}.

According to the SDSS spectroscopic survey (DR1 to DR4), polars make up approximately 20\% of the CVs \citep{inightCatalogueCataclysmicVariables2023}. Another comprehensive survey, conducted by \cite{pala2020}, employing techniques including spectroscopic surveys, variability analysis, X-ray detection, and polarization, scrutinized the volume-limited sample of CVs within 150 pc of the solar neighborhood. They reveal that magnetic WDs constitute 34\% of the total CV sample (including intermediate-polars). However, since most polars do not exhibit outburst activity in optical passbands, the rate of detecting magnetic CVs through optical surveys is only $\sim$1\%. Consequently, only the volume-limited sample is complete. 
 
A possible method to detect polars is to search for their cyclotron radiation, which appears as harmonics of the cyclotron frequency. The contribution from the cyclotron radiation usually manifests as multiple humps in the infrared, optical, or ultraviolet spectra, depending on the magnetic field strength. It also shows phase-dependent variations, especially in the red and infrared passbands. This depends on the inclination of the binary system and the location of the accreted material on the surface of the WD \citep[]{1990cropper,1990schwope,warner95}. 

Overall, polars exhibit distinct observational characteristics due to the high magnetic field of the WD. Their spectra typically show multicomponent emission lines originating from different parts of the system, such as the irradiated secondary star, the ballistic stream, and the magnetic flow. The spectra also include high-ionization lines such as intense HeII (4686 \AA), which is indicative of magnetic accretion \citep[]{1990cropper,2001Worraker}. Cyclotron emission from the accretion region significantly contributes to the system's optical light, with its variability influenced by the changing viewing angle across the orbital period. This variability results in rapid and high-amplitude variations in the orbital light curves of polars, a characteristic rarely observed in nonmagnetic CVs.

Another typical feature that helps differentiate between polars and normal CVs is their X-ray luminosity. Although both magnetic and nonmagnetic CVs can exhibit X-ray radiation, magnetic CVs typically have a higher X-ray-to-optical flux ratio. 
According to \citet{2003ritter}, 11.9\% of magnetic CVs detected with X-ray surveys (using Swift, Chandra, XMM-Newton, and ROSAT) are polars. Therefore, it is possible to identify polars by utilizing their unique X-ray features and cyclotron radiation characteristics specific to normal CVs. All-sky surveys offer valuable insights into the Galactic CV population. For example, recent discoveries of new polars have emerged from the collaboration between SRG/eROSITA and ZTF \citep[]{rodriguezDiscoveryTwoPolars2023,2024galiullin,2024balman,2024schwope}. Additionally, \cite{2022takata} uncovered polar candidates through the ROSAT All-Sky Survey, the second Swift-XRT, XMM-Newton DR10, and TESS. In our prior research aimed at discovering WD binaries within the Galaxy, we integrated \textit{Gaia} Early Data Release 3 (EDR3) and \textit{Zwicky} Transient Facility (ZTF) survey data, successfully  identifying and cataloging over approximately 400 close WD binaries  \citep[CWDBs,][]{2023ren}. Among these CWDBs, we report a source, ZTF J0112+5827, which exhibits an ellipsoidal-like light curve in the $g$ band and two prominent humps in the $i$ and $r$ bands (see their Fig.11), indicating that it might be a polar candidate that deserves further investigation. In this article, we present an extensive, in-depth investigation of ZTF J0112+5827. We explored its potential X-ray emission and conducted time-domain spectroscopic observations of this target. We report that ZTF J0112+5827 is a newly identified polar. The details of our analyses are presented as follows. 

In Sect. 2, we introduce the data-reduction procedures and methods employed in this article. In Sect. 3, we exhibit the main results, including the light curves of ZTF J0112+5827, its phase-resolved spectra, and the Doppler tomograms obtained from these spectra. In Sect. 4, we discuss the science behind our main results. We summarize our conclusions in Sect. 5.  We present all the spectra we observed for ZTF J0112+5827 in the appendix.

\section{Data reduction and methods}

ZTF J01124.48+582757.60 (hereafter ZTF J01124+5827), initially reported by \cite{2023ren}, has been identified as a CV candidate through the ZTF catalog and the \textit{Gaia} EDR3 \citep{2021lindegren}. ZTF J01124+5827 is situated at $\alpha_{\rm J2000}$=01$^h$12$^m$42.44$^s$  and $\delta_{\rm J2000}$=58$^{\circ}$27$'$57.60$''$, with a color index of $G_{\rm BP}$ $-$ $G_{\rm RP}$ = 0.9 mag and an absolute magnitude of $G_{\rm abs}$ = 10.19 mag. This location falls within a typical region for polars.

The light curve of ZTF J0112+5827 is obtained through the public data releases. The ZTF is a wide-field time-domain survey program that utilizes the 48-inch Samuel Oschin telescope at the Palomar Observatory. It can scan the entire northern visible sky (southern sky decl. $>-28^\circ$) at a rate of $\sim3750 \, \text{deg}^2 \, \text{per hour}$ \citep{bellm2019, graham2019, dekanyztf, masci_ztf}.  The instrument operates in three passbands, including $g$ (covering $\sim$4087-5522 $\AA$), $r$ (covering $\sim$5600$-$7316 $\AA$), and $i$ (covering $\sim$6883$-$9009 $\AA$).

\begin{figure*}
\begin{center}
\includegraphics[width=0.75\textwidth]{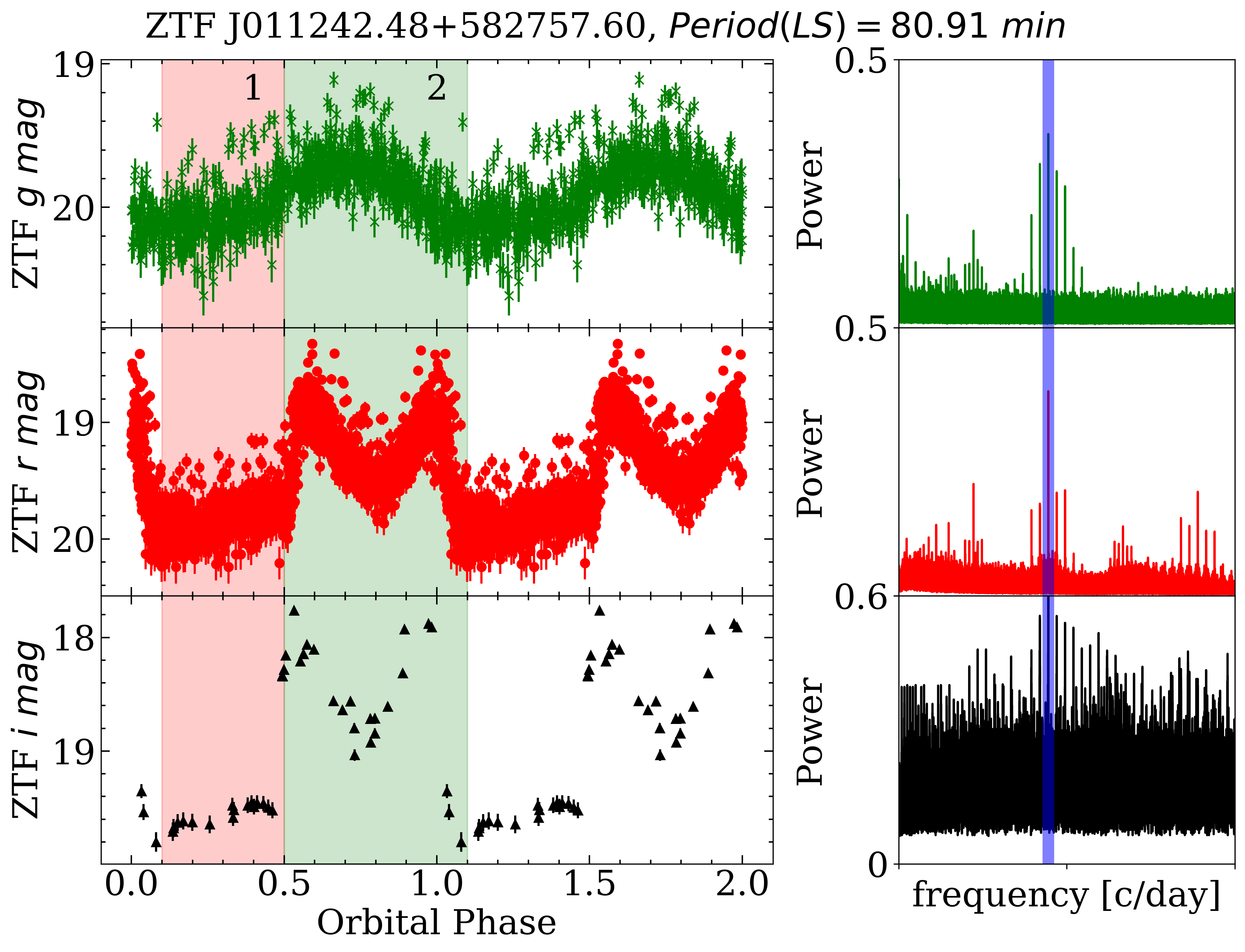}
  \caption{Light curves of ZTF J0112+5827. Left panels: Phase-folded light curves in ZTF $g$-band (green), $r$-band (red) and $i$-band (black) light curves, using the best-fit period of 80.912 minutes. Right panels: Corresponding power spectra from LS periodogram analysis; the aliases around the dominant frequency are separated by a frequency of once per day, reflecting the nightly sampling.}
\label{fig:LS}
\end{center}
\end{figure*}

Since the data from ZTF was updated following \cite{2023ren}, we reconstructed the light curve of ZTF J0112+5827, which is shown in Fig. \ref{fig:LS}. We obtained its latest light-curve data from the ZTF Public Data Release 11 (ZTF DR11). The ZTF observations span from MJD 58400 to MJD 64000, with 798 samples in the $g$ band, 1949 samples in the $r$ band, and 42 samples in the $i$ band. ZTF J0112+5827 has undergone approximately 99,000 orbital cycles. We calculated the period of ZTF J0112+5827 using the Lomb-Scargle periodogram, which, based on the Fourier transform, is employed to detect and characterize periodic components in unevenly sampled time-series data and generate a signal power spectrum. For more information, we invite the reader to consult the Astropy documentation on the Lomb-Scargle method\footnote{\url{https://docs.astropy.org/en/stable/timeseries/lombscargle.html}}.

The phase-folded light curves of ZTF J0112+5827 are shown in Fig. \ref{fig:LS}. The light curves were folded using the following ephemeris:

\begin{equation}
BJD = 2\,458\,263.963(6) + 0.056188(8) \times E
\label{eph}
,\end{equation}where E is the eclipse cycle number, and the initial epoch corresponds to the moment when the Balmer and He II emission lines are at their weakest (see below). The uncertainty of 0.00034 days in the initial epoch is derived from the fundamental timing resolution limited by the ZTF exposure time of 30 seconds. The period uncertainty of 0.0000001 days (0.01 mins) is determined through our period determination algorithm. Both our period measurement and its uncertainty are consistent with \cite{2023ren}. 

To obtain the distribution of radial velocities (RVs) with phase for different components within the ZTF J0112+5827 system, we conducted time-domain spectroscopic observations of ZTF J0112+5827 using the Double Spectrograph \citep[DBSP, ][]{1982oke} on the Palomar 200-inch telescope on October 29 and 30, 2022, with a total observation time of $\sim$15 hrs. The DBSP is a low-to-medium-resolution grating instrument for the Palomar 200-inch telescope, featuring two arms: a blue arm covering a wavelength range of 3400$-$5600\AA\, and a red arm covering a wavelength range of 5650$-$10200\AA. We used a 1.5$\arcsec$ slit, with 158 lines/mm grating in the red arm with a blaze angle centered at 7560$\AA$, and 300 lines/mm grating in the blue arm with a blaze angle centered at 3990$\AA$. Under this configuration, the spectral resolutions are  $\Delta\lambda/\lambda\sim$500 at H$\gamma$ 4340\AA, $\sim$600 at H$\beta$ 4861\AA, and $\sim$400 at H$\alpha$ 6563\AA. We also observed emissions of HeII at 4685\AA\,and HeI at 5877\AA,\, although these emissions are not significant in all phases. These emission lines are used to analyze the structure of the polar system by incorporating them into the Doppler tomography analysis (see Sect. 3.6).

We obtained more than 180 exposures for ZTF J0112+5827, with an average exposure time of 300\,s per observation. Consequently, our observation covered more than 11 periods of the system. Before we observed the target, we conducted an exposure on the standard star BD+28 4211\footnote{\url{https://www.eso.org/sci/observing/tools/standards/spectra/bd28d4211.html}}, with total exposure times of 3\,s for the blue arm and 6\,s for the red arm. The seeing during the observation ranged from $1.0\arcsec$ to $1.2\arcsec$, and the exposures were taken at a line-of-sight elevation reaching up to 63 degrees. The altitude of J0112+5827 was approximately 50 degrees. To further investigate the detailed emission-line profiles, we obtained low-resolution spectroscopic observations using OSIRIS on the 10.4 m Gran Telescopio Canarias on September 14, 2024. Using the R1000B grism with a 0.8$\arcsec$ slit width and 2$\times$2 binning mode, we collected four 480 s exposures covering orbital phases from 0.4 to 0.8, providing enhanced spectral details for line-profile analysis. This instrumental configuration provided spectral resolutions of $\Delta\lambda/\lambda\sim$1000 over a wavelength range of 3630--7880\,\AA. A summary of all data acquired for ZTF0112+5827 is presented in Table \ref{tab:obs_log}.

\begin{table*}
\caption{Observational information for ZTF J0112+5827.}
\label{tab:obs_log}
\centering
\begin{tabular}{lcl}
\hline\hline
Date & Instrument & Specifications \\
\hline
2018-2024 & ZTF & g-band (4087$-$5522 \AA): 798 exposures \\
          &     & r-band (5600$-$7316 \AA): 1949 exposures \\
          &     & i-band (6883$-$9009 \AA): 42 exposures \\
1990-1991 & ROSAT & 0.1$-$2.4 keV X-ray band \\
          &        & 388.50 s exposure \\
29-30 Oct. 2022 & P200/DBSP & Blue: 3400$-$5600 \AA, 1.5 \AA\ resolution, 300 s exp. \\
                &                  & Red: 5650$-$10200 \AA, 1.1 \AA\ resolution, 300 s exp. \\
                &                   & Total of 180+ spectra obtained across 11+ orbital periods \\[0.5ex]
14 Sep. 2024    & GTC/OSIRIS       & Grism R1000B, 3630$-$7880 \AA, \\
                &                   & 4 × 480 s exposures \\
\hline
\end{tabular}
\end{table*}

The spectral data reduction was performed using the standard procedures within the IRAF software package\footnote{\url{https://iraf-community.github.io/}}. Apart from removing cosmic-ray marks and subtracting electronic bias from the images, the sensitivity and nonuniformity of the CCD were reduced using flat-field images. Geometric distortions introduced by the spectrograph's optical system were corrected. The optimal extraction of the target spectrum included background subtraction and wavelength calibration based on He-Ne-Ar lamp frames. Photometric calibration was performed using the images of standard stars.

\begin{table}
\caption{Properties of X-ray source ZTF J0112+5827 as observed by 2RXS.}
\label{tab:data}
\centering
\begin{tabular}{cc}
\hline\hline
Property & Value \\
\hline
name & 2RXS J011242.4+582744 \\
count rate [ct/s] & 0.0623 \\
count rate error [ct/s] & 0.0143 \\
exposure [s] & 388.50 \\
source extent [pixel] & 0.176 \\
source extent prob & 0.29 \\
source quality flag & 0 \\
hardness ratio 1 & 0.505 \\
hardness ratio 2 & -0.273 \\
search offset ["] & 0.223 \\
\hline
\end{tabular}
\end{table}

To depict the high-velocity components in ZTF J0112+5827, we analyzed the variations in the Balmer emissions (H$\alpha$, H$\beta$, and H$\gamma$) and the He II emission line (4685\AA) with the orbital phase using Doppler tomography, which is a technique aimed at reconstructing the distribution of emission regions in two-dimensional velocity space. We used the \texttt{doptomog} code \citep[]{2015kotze,2016kotze}\footnote{\url{https://www.saao.ac.za/~ejk/doptomog/main.html}}, which implements the maximum entropy regularization technique. Flux-modulation Doppler imaging assumes that the variations of emission points over the orbital period have a sinusoidal form. Each point in this space is described by two polar coordinates: $v$ and $\theta$. Here, $v$ is the absolute velocity value projected along the line of sight relatively to the center of mass, and $\theta$ is the angle between the velocity vector and the direction from the center of mass to the donor. Doppler tomograms can be reconstructed using two types of projections: standard projection and inside-out projection. The standard projection is suitable for studying low-velocity structures, such as the radiation hemisphere of the donor or the accretion stream segment near the Lagrange point L1. When gas around the WD accelerates to high speeds, traces on the Doppler maps in standard projection tend to spread out over a larger area, appearing blurred. For analyzing high-velocity structures concentrated in more compact regions with higher contrast, the inside-out projection is preferred \citep[]{2015kotze,2016kotze}.

The Doppler maps were constructed using the input spectra obtained from the P200 telescope \href{https://zenodo.org/records/14591501/files/spectraP200.png}{(supplementary material)}. To facilitate the interpretation and analysis of the Doppler map for ZTF J0112+5827, we overlaid binary models onto the map. It is crucial to note that the mass of the WD in magnetic CVs is generally higher than in nonmagnetic CVs. Specifically, magnetic WDs have an average mass of $0.784 \pm 0.047 M_\odot$, whereas nonmagnetic WDs typically have a mass of $0.663 \pm 0.136 M_\odot$ \citep{2015ferrario}. Taking this into account, along with the uncertain mass of the donor star, we assumed a WD mass of $m_1 = 0.80 M_\odot$ for our model. Based on the evolutionary-period distribution for CVs, the donor star's mass was set at $m_2 = 0.074 M_\odot$ \citep{knigge2011, 2023scaringi}. The absence of eclipses, alongside the observation of the companion star partially blocking the accretion stream, led us to the assumption of an inclination angle of 70 degrees for the Doppler tomogram.

\begin{figure}
     \centering
     \includegraphics[width=0.9\linewidth]{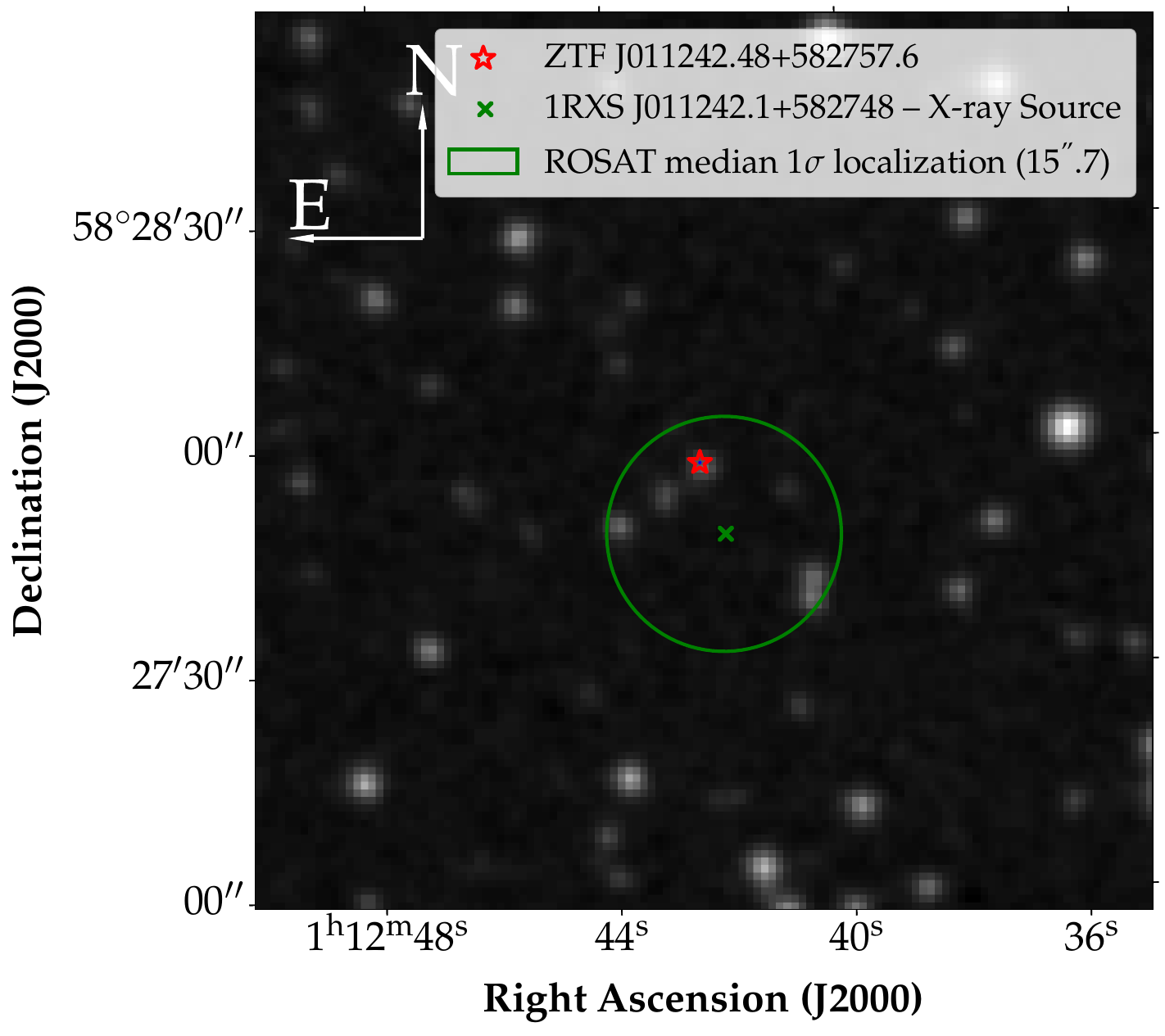}
     \caption{$2\arcmin \times 2\arcmin$ POSS2/UKSTU Red survey image of ZTF J2243+5242. The red asterisk marks the source labeled ZTF J01124+5827. The ROSAT median $1\sigma$ localization is depicted with a green uncertainty circle of 15.7 arcseconds. }
     \label{fig:xray_image}
 \end{figure}

We retrieved the X-ray data for ZTF J0112+5827 using The ROentgen SATellite (ROSAT). The ROSAT was designed to perform a comprehensive all-sky X-ray survey with high sensitivity in the energy range of 0.1–2.4 keV \citep{1986trumper}. The ROSAT 1RXS catalog, which includes both a bright and faint source catalog, contains approximately 127,730 sources. Subsequently, the ROSAT point-source catalog was updated and renamed as the Second ROSAT All-Sky Survey Point Source Catalog (2RXS), comprising 135,000 X-ray sources \citep{2016rosat3} with a flux limit of approximately $2 \times 10^{-13}$ erg s$^{-1}$ cm$^{-2}$. We matched ZTF J0112+5827 with 2RXS at 15.7 arcseconds (corresponding to the median 1$\sigma$ localization), as shown in Fig. \ref{fig:xray_image}. The results of this search are presented in Table \ref{tab:data}. The observed X-ray flux of ZTF J0112+5827 in the 0.1–2.4 keV energy band is reported as $(68.4 \pm 15.7) \times 10^{-14}$ erg s$^{-1}$ cm$^{-2}$.
  
\section{Main results}

\subsection{Light curves}

Figure \ref{fig:LS} exhibits the light curves of ZTF J0112+5827 in the $g$,$r$, and $i$ passbands. All these results are fully consistent with \cite{2023ren} (see their Fig. 11). The zero phase we selected corresponds to the moment when the Balmer and helium emission lines are at their weakest in the spectrum, which we attribute to the companion star being closest to us, thereby obscuring the emission structure of the entire binary system. ZTF J0112+5827 exhibits an ellipsoidal-like variability curve in the $g$ band. However, two prominent humps emerge when observing its light curves in the $i$ and $r$ bands. The double humps are located at phases $\sim$0.63 and $\sim$0.97 and are symmetric around phase $\sim$0.82, where the flux comes to a local minimum.

Based on the variability characteristics of ZTF J0112+5827, we divided its phases into two regions. We refer to orbital phases from 0.1 to 0.5 as Region 1, and the rest part (from 0.5 to next 0.1) as Region 2, which is indicated in Fig. \ref{fig:LS}. The typical features of these two regions are described below:

\begin{itemize}
\item Region 1: In both bands, we observe a relatively stable flux. The light curves exhibit a gradual increase trend, with a brightness increase of approximately 0.2 to 0.3. 

\item Region 2: The light curve exhibits a double-peaked structure around phases 0.63 and 0.97 in the $i$ and $r$ bands, while the peak brightness in the $i$ band can vary by 1.2 mag. In the $i$ band, the peak brightness can vary by up to 2 mag.

\end{itemize}

\subsection{Multi-phase spectra}

\begin{figure*}[htpb]
\begin{center}
\includegraphics[width=1\textwidth]{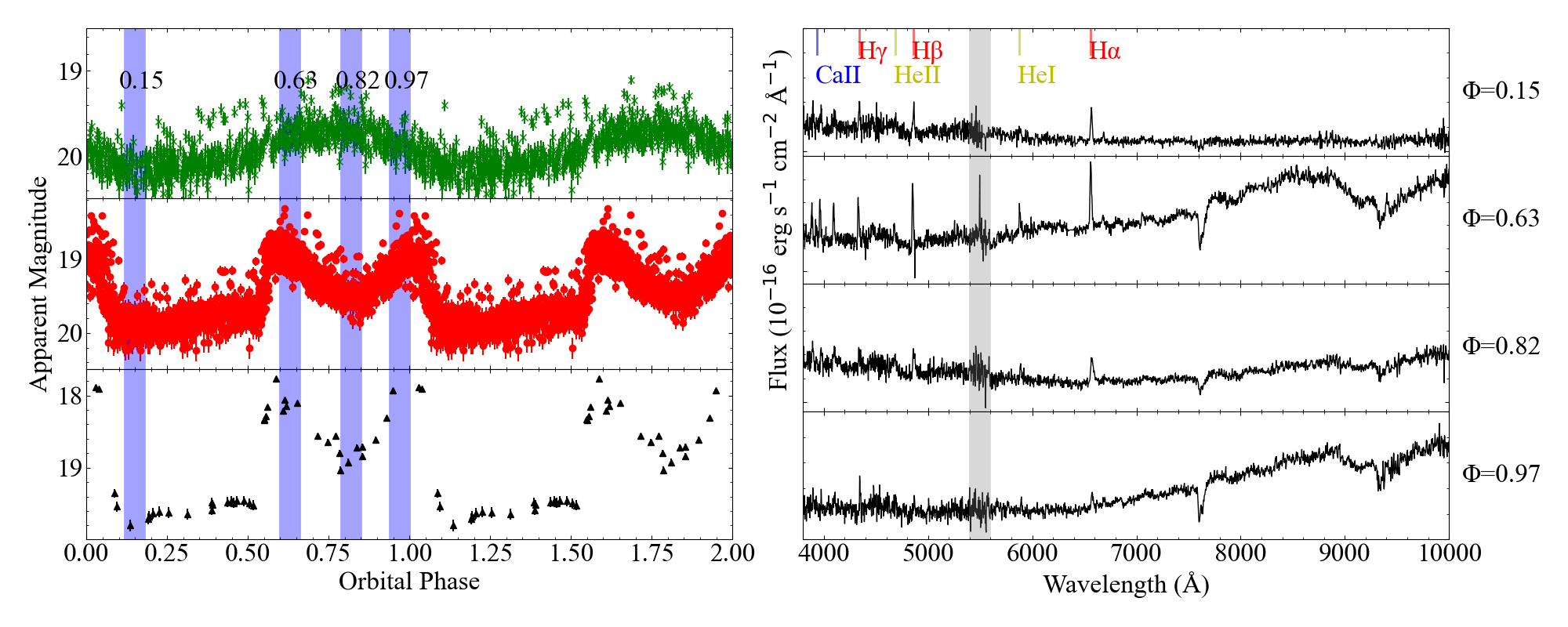}
\caption{ZTF photometry taken over two orbits (left) and multi-phase spectroscopy (right) of ZTF J0112+5827 are presented. The blue highlighting in the left panels marks the phases of the spectra. The right panels display example spectra corresponding to phases of 0.15, 0.63, 0.82, and 0.97. The gray areas indicate the regions where red and blue spectra are stitched together, accompanied by a relatively low signal-to-noise ratio.}
\label{fig:spectrum}
\end{center}
\end{figure*}

We combined ZTF photometric data ($g$, $r$, $i$ bands; MJD$>$59000) with our spectroscopic observations (MJD 59883$-$59884) for analysis, which are shown in Fig. \ref{fig:spectrum}. We calculated the phases using the ephemeris Eq. \ref{eph}. The detailed spectra are available in \href{https://zenodo.org/records/14591501/files/spectraP200.png}{supplementary material}.

As expected, we observed significant spectral variations in ZTF J0112+5827 with respect to the phase. The spectral energy distribution (SED) shows a more pronounced shift toward the red portion compared to the blue region. This is notably highlighted by a distinct increase in the SED around 5500\AA\, toward the red end after phase 0.63. In terms of spectral lines, the predominant features stem from the Balmer series emission lines, specifically H$\alpha$, H$\beta$, and H$\gamma$. Additionally, there are discernible emission-line characteristics of He I at 5877\AA\, and He II at 4685\AA, albeit with a lower signal-to-noise ratio (S/N) compared to the Balmer lines. The relative intensities of these emission lines vary with the phase, peaking at around phase 0.63, declining overall after phase 0.82, reaching a minimum near phase 0.97, and then gradually rising once more. The phase variations of the Balmer emission lines, observed at low resolution, are depicted in the supplementary material available at \href{https://zenodo.org/records/14591501/files/Emission_line.png}{Supplementary Material}. In the lower section of Fig. \ref{fig:spectrum}, we present four sample spectra, with their exposure midpoints corresponding to phases of 0.15, 0.63, 0.82, and 0.97, respectively. We also observed some less continuous features, around phase 0.63; these emission lines displayed possible splitting characteristics, with the split parts exhibiting opposite redshifts and blueshifts. However, due to the relatively low SNR and spectral resolution of our observations, these features were not always discernible. 

\begin{figure*}[htpb]
\begin{center}
\includegraphics[width=0.75\textwidth]{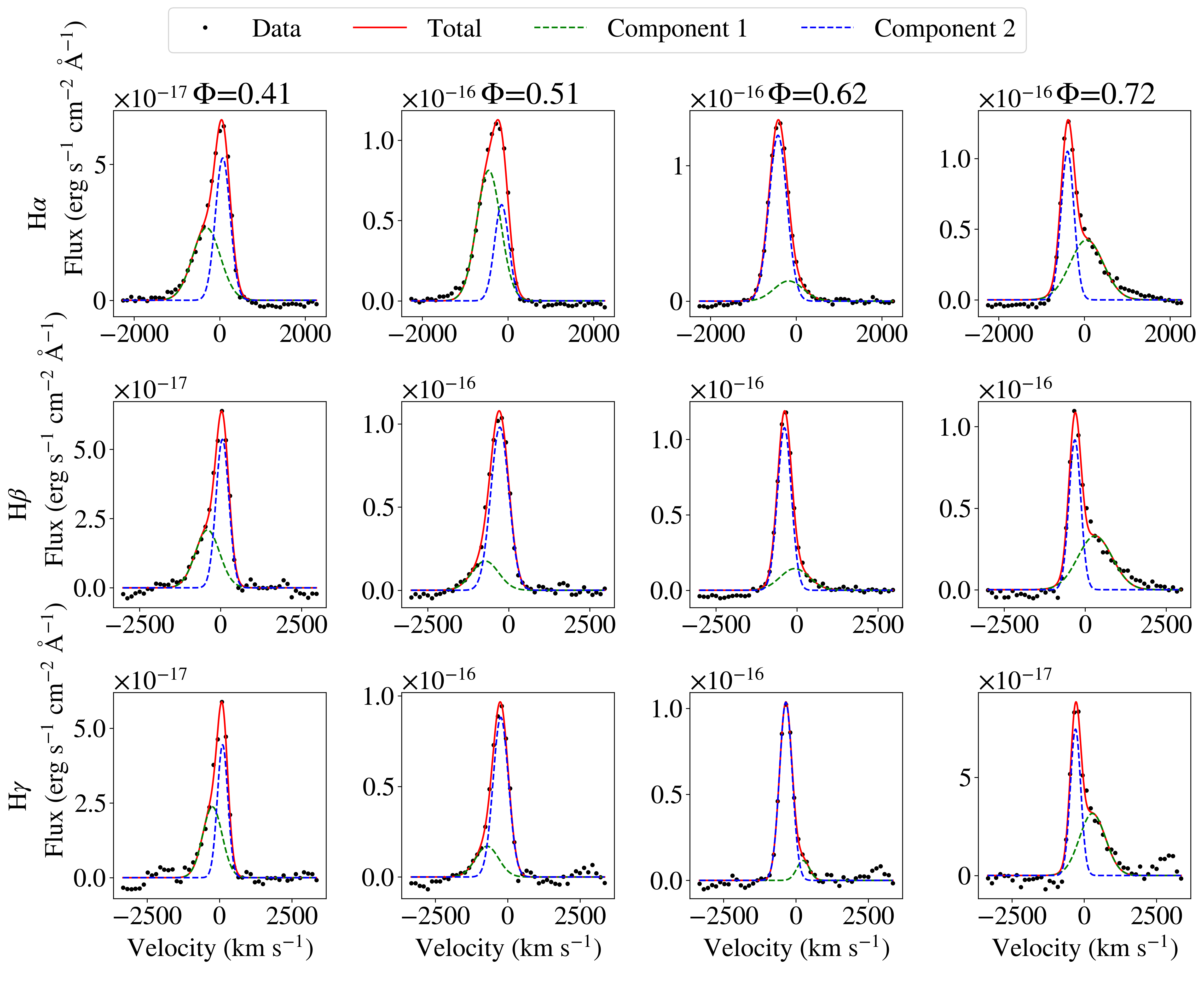}
\caption{Spectral decomposition of the three Balmer lines for ZTF J0112+5827 at four different orbital phases. Each row represents a different phase, and each column shows a different emission line. The black points show the observed spectra, and the solid red lines represent the total fit. The blue and green dashed lines represent the two fit Gaussian components.}
\label{fig:Emission_line_fit}
\end{center}
\end{figure*}

To investigate the detailed emission line profiles of ZTF J0112+5827, we analyzed the OSIRIS/GTC spectra by decomposing each emission line into a superposition of two Gaussian components as shown in Fig. \ref{fig:Emission_line_fit}. At phase 0.41, a broad component is primarily blueshifted. At phase 0.51, two components are mixed. From phases 0.62 to 0.72, a broad component becomes redshifted and increases in flux.


\subsection{Doppler tomography}

\begin{figure*}
\begin{center}
\includegraphics[width=0.9\textwidth]{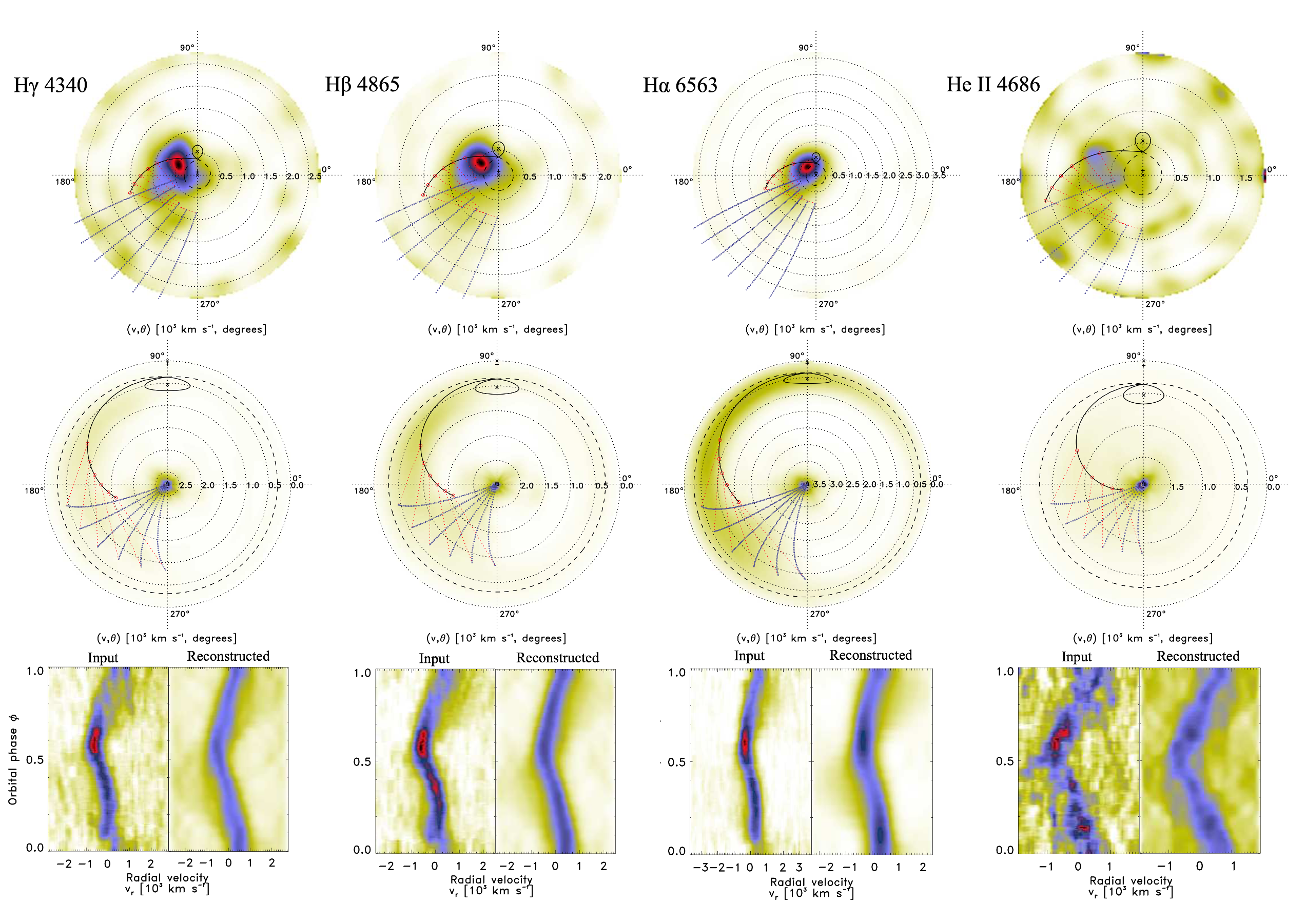}
\caption{Doppler tomograms of Balmer emissions (H$\alpha$, H$\beta$, and H$\gamma$) and He II emission, along with their trailed spectra. The upper panel shows the standard Doppler tomograms, the middle panel presents the inside-out projections, and the bottom panel shows the trailed input and reconstructed spectra. The velocities of the Roche lobes of the WD (shown as the black dashed closed line) and the donor star (shown as the black continuous closed line) components are overlayed on the tomograms. The velocities of the ballistic trajectory (shown as the black continuous line) and the magnetic part of the trajectory (shown as blue dashed lines) are superimposed.}
\label{fig:dop}
\end{center}
\end{figure*}

Figure \ref{fig:dop} presents the Doppler tomograms of the Balmer emission line (H$\alpha$, H$\beta$, and H$\gamma$), He II, along with their trailed spectra. The He I 5875.6 \AA{} emission is not utilized due to strong contamination from Na I 5895.9 \AA{}. Based on the binary parameters and the Doppler tomograms of the emission line, we overlaid binary models onto the Doppler map. This overlay includes the Roche lobe of the primary star (indicated by a dashed elliptical line), the single-particle ballistic stream trajectory from the secondary star to the primary star (solid line), the magnetic dipole field lines around the primary star (depicted as thin, dashed blue lines), and the Roche lobe of the secondary star (also indicated by a dashed elliptical line).

In the upper panel of Fig. \ref{fig:dop}, the standard projection tomograms of the Balmer emissions (H$\alpha$, H$\beta$, and H$\gamma$) and He II emission are displayed. All Doppler tomograms of Balmer emissions reveal a highly dispersed region. In the binary models, the irradiated secondary star combines with the ballistic stream. This excessively compacted blend of emissions creates an intensely bright spot within the velocity range of 250$-$700 km s\(^{-1}\), $90-180^\circ$. Meanwhile, we observe a fainter and more diffuse emission along the model's magnetic dipole field lines within the velocity range of 0$-$2000 km s\(^{-1}\), $180-270^\circ$. The He II 4686\AA{} tomogram also exhibits similar features. However, it is very noisy due to its low SNR. 

In the middle panel of Fig. \ref{fig:dop}, The inside-out projections more effectively separate the low-velocity emission components and reveal the high-velocity components produced as the magnetically confined accretion stream falls toward the primary star. These low-velocity components (0$-$1000 km s$^{-1}$, $90-270^\circ$) form a brighter ridge along the model stream trajectory. The high-velocity components (0$-$2500 km s$^{-1}$, $180-270^\circ$) are distributed along the model's magnetic dipole field lines.

In the bottom panel, the trailed spectra (both input and reconstructed) exhibit sinusoidal-like patterns, with amplitude velocities of approximately 450 km s$^{-1}$. There is a significant presence of strongly redshifted velocity components with weaker emission between phases 0.5 and 1.

\section{Discussion}

\subsection{The magnetic field of ZTF J0112+5827}

The relation between X-ray flux ($F_X$) and optical flux ($F_{\text{opt}}$) can serve as an effective diagnostic tool to differentiate magnetic CVs, including intermediate polar and polar one, from nonmagnetic CVs such as dwarf novae and nova-like systems, as well as to characterize them in different states \citep[quiescent and outburst and see][]{mukai2017,2009agueros}. For polars, the presence of a strong magnetic field in the WD leads to direct control of the accretion flow from the companion star to the WD along the magnetic field lines, avoiding the need for a standard accretion disk structure. This direct magnetic field control results in intense X-ray emission when matter impacts the WD surface, while optical emission is relatively weaker due to the absence of a standard accretion disk structure.

\begin{figure}[ht!]
\begin{center}
\includegraphics[width=0.45\textwidth]{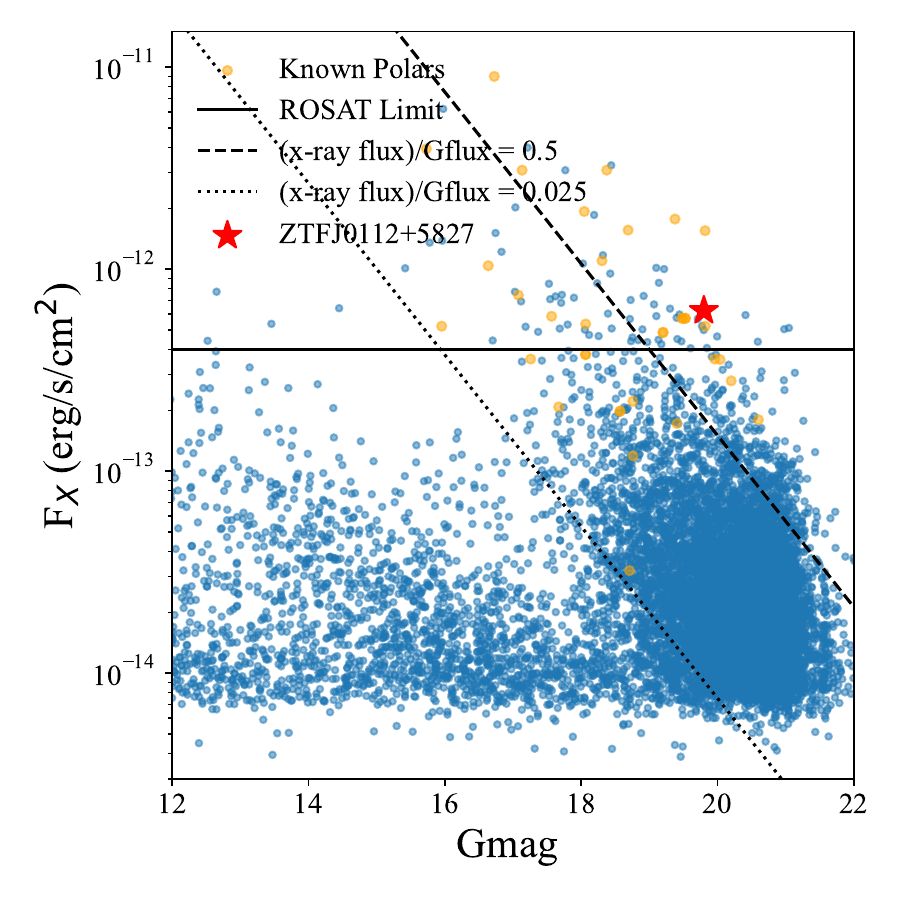}
  \caption{Correlation of \textit{Gaia} G-band magnitudes of sources with their X-ray fluxes measured in units of mW/m$^2$. All SRG/eFEDS Galactic objects are represented as blue circles, the majority of which fall below the detection threshold of the ROSAT All Sky X-ray survey, which is indicated by the solid black line. The dashed and dotted black lines represent the X-ray-to-optical flux ratios of 0.5 and 0.025. The newly identified polar ZTF J0112+5827 is denoted by the red asterisk. Previously known polars \cite{2020Abril, inightCatalogueCataclysmicVariables2023,rodriguezDiscoveryTwoPolars2023} are shown as orange dots. All confirmed polars exhibit characteristically high X-ray-to-optical flux ratios, with most sources having (F$_X$/F$_{opt}$) > 0.025, and a significant fraction exceeding (F$_X$/F$_{opt}$) > 0.5.
  }
\label{fig:X_ray}
\end{center}
\end{figure}

The X-ray-to-optical flux ratio ($F_X/F_{\text{\rm opt}}$) in polars is higher than in other types of CVs. Figure \ref{fig:X_ray} presents ZTF J0112+5827 along with other reported polars (with the catalog sourced from  \cite{2020Abril}, \cite{inightCatalogueCataclysmicVariables2023} and \cite{rodriguezDiscoveryTwoPolars2023}). This figure compares the X-ray flux with the optical flux from \textit{Gaia} DR3. We find that all confirmed polars demonstrate characteristically high X-ray-to-optical flux ratios, with the majority exhibiting (F$_X$/F$_{\rm opt}$) > 0.025, and a substantial fraction surpassing (F$_X$/F$_{\rm opt}$) > 0.5. Notably, ZTF J0112+5827 is positioned above the (F$_X$/F$_{\rm opt}$) = 0.5 threshold, suggesting its emission is dominated by X-ray radiation. This strong X-ray flux of ZTF J0112+5827 indicates that it is possibly a polar system.

 In the accretion process occurring on the magnetic WD, electrons within the accretion region generate cyclotron radiation, with the predominant contribution to the flux arising in the optical and near-infrared wavelengths \citep{gansicke2001,2016fuchs}. As shown in Fig.\,\ref{fig:LS}, the light curve of ZTF J0112+5827 shows two spikes around phases $\sim$0.63 and $\sim$0.97 in the $i$ and $r$ bands. Moreover, the spectra exhibit cyclotron radiation humps \citep[]{1990cropper,2000wickramasinghe,littlefield2017}. In  Fig.\,\ref{fig:LS}, around phase 0.63 of the spectrum of ZTF J01124+5827, following the peak brightness in the $i$ and $g$ band, the cyclotron humps manifest in the reddest region of the spectrum. The double-peaked structure is frequently observed in the light curves of polars, and the cyclotron radiation reaches its maximum flux when the line of sight is approximately $90 ^\circ$ to the magnetic field \citep[]{1990cropper,2000wickramasinghe,gansicke2001}.

Common methods for measuring the magnetic field on the surface of WDs include the following
\citep{2000wickramasinghe,1990cropper}:
(1) Zeeman spectroscopy, which includes analyzing the splitting and polarization of spectral lines due to the Zeeman effect, allowing us to infer the local magnetic-field strength and orientation;
(2) continuum polarimetry, which involves measuring the linear and circular polarization of the continuum radiation to estimate the field strength; and
(3) cyclotron radiation. In this method, the magnetic-field strength is determined by fitting the observed SED with a model that incorporates cyclotron harmonics.

For ZTF J0112+5827, due to the lack of polarization observations, we can only calculate its magnetic-field strength by fitting its SED with a cyclotron harmonics model. To accomplish this, the viewing angle $\theta$ of the cyclotron beaming, magnetic-field structure, and the field orientation become crucial parameters in the modeling procedures \citep{2015ferrario,2019mason}.

According to \cite{1969Ramaty}, under conditions of rapid Faraday rotation, the transfer equations for polarized radiation are simplified and reduced to independent equations for the ordinary and extraordinary waves. The cyclotron radiation intensity for the two modes can be expressed as
\begin{equation}
I_{\theta \pm}=I_{R J}\left(1-\exp \left(-\alpha_{\theta \pm} \Lambda\right)\right),
\end{equation}
where $I_{R J}=k_B T_e \omega^2 / 8 \pi^3 c^2$ is the Rayleigh-Jeans intensity per polarization mode ($k_B$ is Boltzmann's constant) and $\alpha_{\theta \pm}$ are cyclotron absorption coefficients in $\omega_p^2 / \omega_c c$ units. The coefficients $\alpha_{o, e}$ in this work were calculated according to the method.

The absorption coefficients $\alpha_{\pm}$ are expressed in units of $\omega_p^2/\omega_c c$, where $\omega_c = \frac{eB}{m_e c}$ is the cyclotron frequency, $\omega_p = \sqrt{\frac{4\pi N_e e^2}{m_e}}$ is the plasma frequency, and $N_e$ is the electron density. The dimensionless plasma parameter (size parameter) $\Lambda$ is determined by

\begin{equation}
\Lambda = \frac{\omega_p^2}{\omega_c c}\ell
,\end{equation}
where $\ell$ represents the depth of the radiating region along the line of sight. The total radiation intensity $I$ is calculated as the sum of the ordinary and extraordinary wave intensities:

\begin{equation}
I = I_{\theta+}+ I_{\theta-}
.\end{equation}

Simulating cyclotron radiation requires determining the absorption coefficient $\alpha_{\pm}$. This is achieved by convolving the emission coefficient of a single electron with a relativistic Maxwellian distribution, assuming the medium is in thermodynamic equilibrium (satisfying Kirchhoff's law) \citep[]{1980chan,2020kolbin,2022kolbin}. 

Cyclotron radiation simulation requires multiple parameters including  magnetic-field strength ($B$), electron temperature ($T$), viewing angle ($\Theta$), shock structures, and so on. To simplify the calculation, we adopted the constant-lambda (CL) model, which assumes radiation emerges from a single path through a slab of uniform optical-depth parameter $\Lambda$. We employed the CL code originally developed by \citet{1990schwope} and used by 
 \citet{2008campbella} to produce model cyclotron spectra. CL cyclotron models depend on four distinct global variables: $B$, $T_{e}$, $\Theta$, and $\Lambda$. Variations in these parameters lead to complex and quasi-degenerate changes in the cyclotron spectra. 
We selected the spectrum corresponding to the peak of the $r$-band light curve and assumed its $\Theta = 90^{\circ}$. Since $\Lambda$ only affects the amplitude of the flux without altering the spectral overall shape, the cyclotron radiation spectrum is normalized, and we can disregard fitting to different values of $\Lambda$.

We selected the 0.63 phase (approximating an angle of $90^{\circ}$) and subtracted the 0.15 phase (which has no cyclotron radiation flux). We then normalized the spectral flux. The best-fit parameters for the magnetic-field strength ($B$) and electron temperature ($T_{\mathrm{e}}$) were determined by minimizing the $\chi^2$ statistic:

\begin{equation}
{\chi^2}(B, T_{\mathrm{e}})=\sum_{i=1}^n \left(\frac{F_{\mathrm{obs},i} - F_{\mathrm{mod},i}(B,T_{\mathrm{e}})}{\sigma_i}\right)^2,
\end{equation}
where $F_{\mathrm{obs},i}$ and $F_{\mathrm{mod},i}(B,T_{\mathrm{e}})$ are the observed and model-normalized fluxes at each wavelength point $i$, respectively, and $\sigma_i$ represents the corresponding observational uncertainty.

\begin{figure}[ht]
\begin{center}
\includegraphics[width=0.45\textwidth]{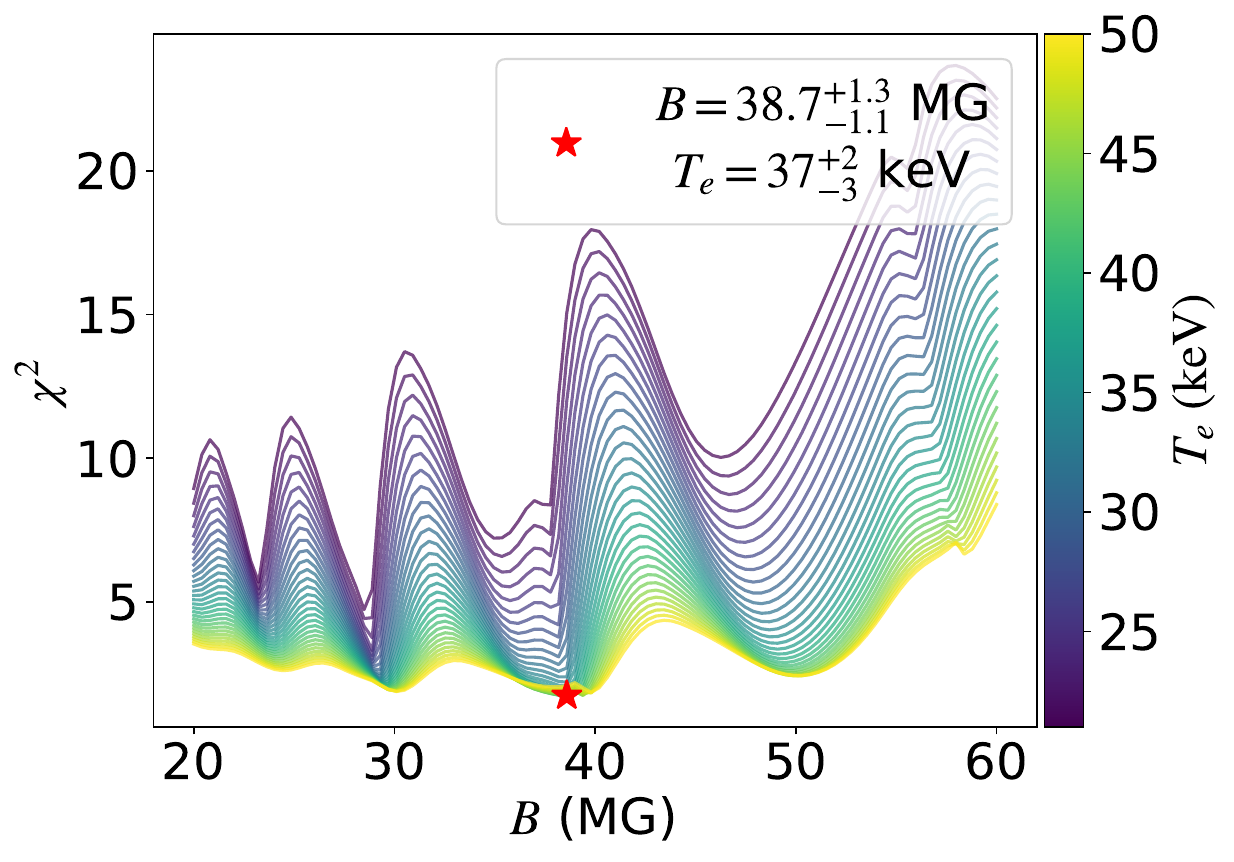}
\caption{$\chi^2$ distribution for cyclotron spectrum fitting of ZTF J0112+5827. The color scale represents different temperatures from 20 to 50 keV. The magnetic field strength varies from 20 to 60 MG. The best-fit parameters are $B = 38.7_{-1.1}^{+1.3}$ and $T_e = 37_{-3}^{+2}$ keV (marked by red star), which reproduce the observed cyclotron harmonics in the optical spectra well.}
\label{fig:kafan}
\end{center}
\end{figure}

\begin{figure}[ht]
\begin{center}
\includegraphics[width=0.5\textwidth]{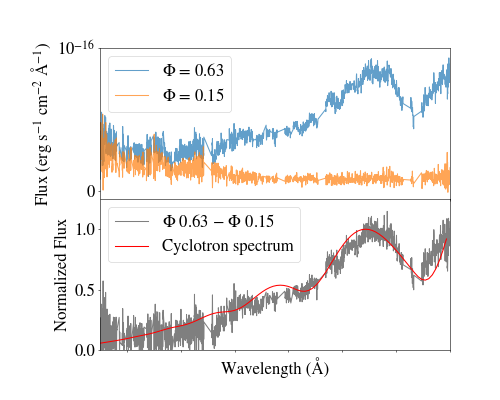}
\caption{Cyclotron spectrum analysis of ZTF J0112+5827. Top panel: Original spectra at phase 0.63 and phase 0.15 after removing emission lines and telluric features. Bottom panel: Difference spectrum (black line) was obtained by subtracting the phase 0.15 spectrum from the phase 0.63 spectrum, showing clear cyclotron harmonics. The best-fit theoretical cyclotron spectrum (red line) corresponds to a magnetic-field strength of $B = 38.7_{-1.1}^{+1.3}$ MG and electron temperature of $T_e = 37_{-3}^{+2}$ keV.}
\label{fig:best}
\end{center}
\end{figure}

We performed spectral fitting across a parameter space spanning $T_e$ from 20 to 50 keV and $B$ ranging from 20 to 60 MG. The optimal parameters were determined through $\chi^2$ minimization, as illustrated in Fig. \ref{fig:kafan}. We determined the 99$\%$ confidence intervals of the $\chi^2$ statistic to assess the uncertainties (see, e.g., \cite{2007nr}). The revised analysis reveals three possible solutions for the magnetic field strength, with the following best-fit values and uncertainties:

\begin{itemize}
    \item $B = 29.4_{-3.1}^{+2.7} \, \mathrm{MG}$ with a minimum $\chi^2$ value of 1.8764
    \item $B = 38.7_{-1.1}^{+1.3} \, \mathrm{MG}$ (global minimum) with a minimum $\chi^2$ value of 1.6379
    \item $B = 49.6_{-2.1}^{+1.7} \, \mathrm{MG}$ with a minimum $\chi^2$ value of 2.4226
\end{itemize}
Among these, the solution with $B = 38.7_{-1.1}^{+1.3} \, \mathrm{MG}$ provides the global minimum and is considered the most representative. The other possible solutions at $B = 30.1_{-3.1}^{+2.7} \, \mathrm{MG}$ and $B = 49.9_{-2.1}^{+1.7} \, \mathrm{MG}$ are also worth noting due to the complexity of the parameter space. A comparison between the best-fit model spectrum for $B = 38.7_{-1.1}^{+1.3} \, \mathrm{MG}$ and the observed data, including contributions solely from the accretion regions, is presented in Fig. \ref{fig:best}.

In summary, ZTF J01124+5827 exhibits stronger emission in the X-ray band compared to the optical band. Its light curves and spectra both display characteristics indicative of cyclotron radiation. The spectrum also shows high-ionization lines such as the prominent He II $\lambda$ 4686 \AA line. All these features support the idea that it is a polar system. We have determined its magnetic field strength to be $B = 38.7_{-1.1}^{+1.3}$ MG.

\subsection{The accretion structure of ZTF J0112+5827}
\begin{figure}[ht]
\begin{center}
\includegraphics[width=0.45\textwidth]{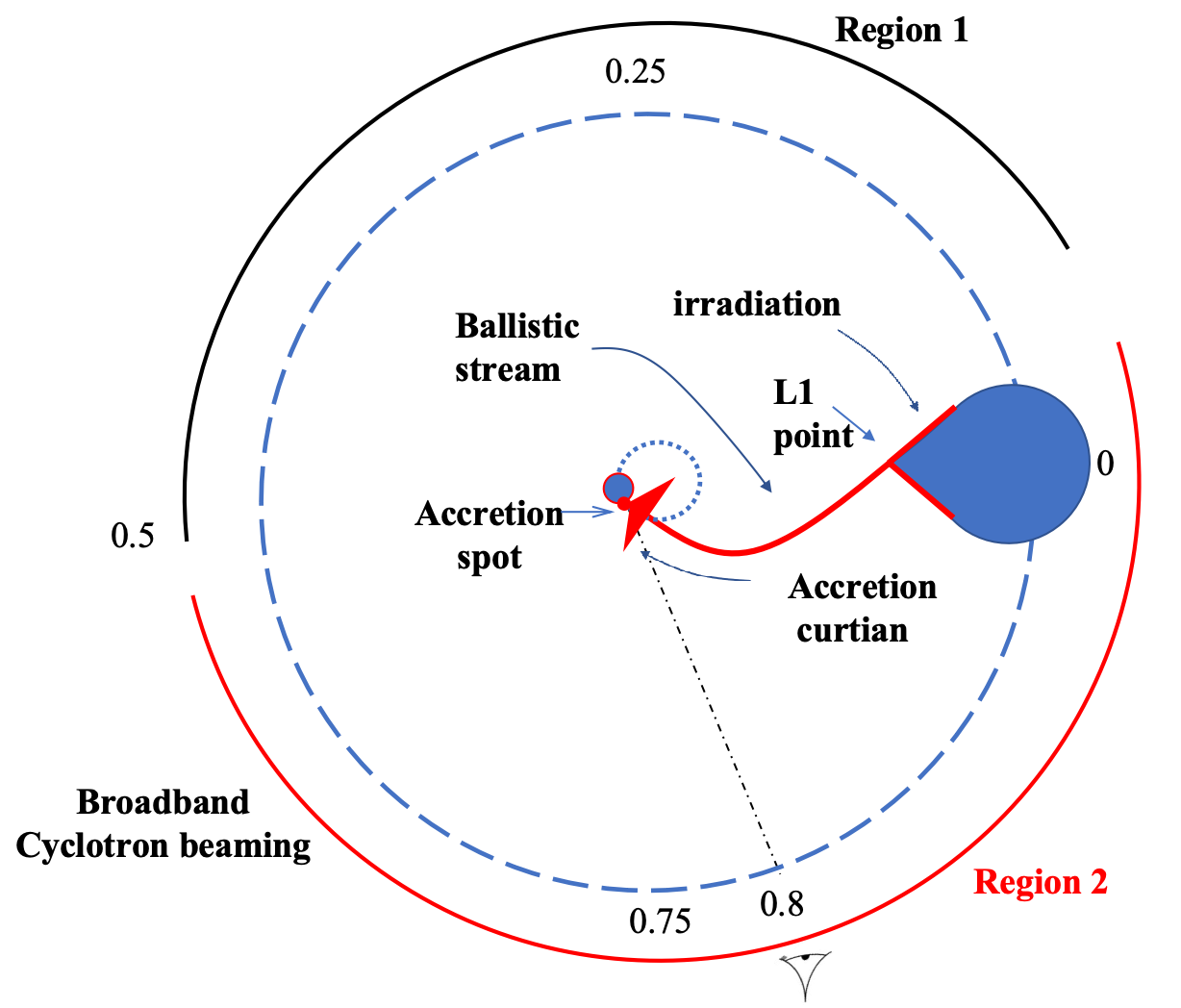}
  \caption{Cartoon of the orbit of ZTF J0112+5827 illustrating the observer's viewpoint as rotating along the dashed circle relatively to the orbital phase. At phase 0, the donor partially obscures the stream. Cyclotron radiation begins at phase 0.5. By phase 0.8, the accretion spot is most closely aligned with the observer's line of sight.}
\label{fig:cartron}
\end{center}
\end{figure}

We analyzed data from various phases to elucidate the accretion structure of ZTF J0112+5827. The light curve displays a secondary minimum (at phase $\sim$0.82) between two peaks when the angle $\theta$ between the accretion spot and the observer's line of sight reaches its minimum value \citep[]{gansicke2001,2020kolbin}. From the standard tomograms of Fig. \ref{fig:dop}, we can see that ZTF J0112+5827 does not have an accretion-disk structure. From standard Doppler tomograms, we observe a strong, intensive bright spot (250$-$700 km s\(^{-1}\), $90-180^\circ$) that dominates the brightness of the Doppler map. When combined with the binary models, this suggests that the emission mainly originates from the ballistic stream. Additionally, emissions are seen distributed along the magnetic dipole field lines of the model (0$-$2000 km s\(^{-1}\), $180-270^\circ$), originating from the accretion curtain near the WD's surface. In the inside-out projections, we can clearly distinguish between the low-velocity emissions from the ballistic stream and the high-velocity emissions from the magnetically confined accretion flow. The low-velocity emission (0$-$200 km s\(^{-1}\), $90-180^\circ$) is located in the model's secondary star's Roche lobe and along the stream trajectory. The funneling of the emission (0$-$2000 km s\(^{-1}\), $180-270^\circ$) follows the model's magnetic dipole trajectories.

In Fig. \ref{fig:dop}, the trailed spectra of the Balmer emission lines (H$\alpha$, H$\beta$, and H$\gamma$) and the He II emission line reveal the main components observed at different phases. At phase 0$-$0.08, the Balmer emissions are at their weakest, suggesting that the donor star obscures a substantial portion of the ballistic stream \citep[]{2016fuchs,2022kolbin}, resulting in a diminished strength of the emission lines. Consequently, we designate this as phase 0. The trailed spectra of H$\gamma$ and H$\beta$ also reveal a high-velocity redshifted component, likely emanating from the accretion curtain.

At an orbital phase of $\sim$0.1, the ballistic flow begins to emerge, and as the phase increases from $\sim$0.1 to $\sim$0.5, the flux in all three bands increases. According to the study by \citet{2016fuchs}, a gradually increasing trend in flux in all three bands is not observed in the low state (characterized by low accretion rates and the absence of strong synchrotron radiation), but it is evident in the high state. This may be attributed to the increased accretion rate and heating in the high state, where the bright accretion stream enhances the overall flux. We propose that the higher accretion rate in the high state heats the accretion flow, and as the phase increases, the obstruction by the accretion curtain decreases, leading to an increase in overall flux.

As previously presented, we observe that the Balmer emission lines exhibit two potential components: a primary emission and a relatively weaker secondary emission, with their relative velocities varying with phase. Based on the trailed spectra, we infer that the primary emission arises from the accretion stream, while the secondary emission originates from the accretion curtain near the surface of the WD. Between phase $\sim$0.1 and 0.4, the stream possesses a greater positive line-of-sight velocity compared to the curtain, resulting in the observed redshift of the primary emission relative to the secondary one. Conversely, at an orbital phase of 0.6, when the accretion stream's velocity is directed toward our line of sight, these lines achieve their maximum blueshift. Between phases $\sim$0.6 and $\sim$0.9, weaker, relatively redder emission lines are observed, primarily due to the presence of the accretion curtain. After phase 0.9, the occlusion of the donor star became evident, and all emission lines began to diminish significantly.
To illustrate the phenomena observed at various phases of ZTF J01124.48+582757.60, we  constructed a schematic representation of its orbital motion, as depicted in Fig. \ref{fig:cartron}.

\subsection{ZTF J0112+5827: A potential period-bounce system}

We can also deduce whether ZTF J0112+5827 has undergone period bounce by estimating its accretion rate. The observed X-ray flux of ZTF J0112+5827 in the 0.1–2.4 keV energy band is $F_{0.1-2.4 \mathrm{keV}} = 6.84 \pm 1.57 \times 10^{-13} \mathrm{erg\,cm}^{-2} \mathrm{~s}^{-1}$. Based on \textit{Gaia} distance constraints, the X-ray luminosity is $L_{0.1-2.4 \mathrm{keV}} = 1.08 \pm 0.24 \times 10^{31} \mathrm{erg\,s}^{-1}$. We applied the bolometric correction factor from \citet{2024galiullin}, approximately in the range of 3.3–14.8, to estimate the luminosity in the 8.5-100 keV band, although this value might be somewhat inaccurate. We estimated the ratio $R_{\mathrm{WD}} / M_{\mathrm{WD}}$ based on a maximum velocity of 2000 km/s from the inverse Doppler diagram, using the formula $v = \sqrt{\frac{2 G M_{\mathrm{WD}}}{R_{\mathrm{WD}}}}$. The resulting accretion rate is calculated as 
\begin{equation}
    \dot{M} = \frac{2 L_{\mathrm{X}} R_{\mathrm{WD}}}{M_{\mathrm{WD}} G} \approx (1.4-6.3) \times 10^{-11} M_{\odot}\mathrm{yr}^{-1}.
\end{equation}
When the binary system reaches the minimal period of 78 minutes, the companion star's mass drops to approximately 0.06 $M_\odot$ and continues to decrease as the binary period lengthens. By the time the period extends to 80 minutes, the mass is about 0.05 $M_\odot$. This evolution process slows down, and the mass transfer rate sharply declines to a level of around $10^{-12} M_{\odot}\,\mathrm{yr}^{-1}$, rendering the binary system faint and difficult to detect. However, the estimated higher accretion rate for ZTF J0112+5827 suggests that its companion star retains a relatively higher mass, indicating that ZTF J0112+5827 is a polar system that has not yet experienced a period bounce.

\subsection{Gravitational-wave intensity}

\begin{figure}[htpb]
    \centering
    \includegraphics[width=1\linewidth]{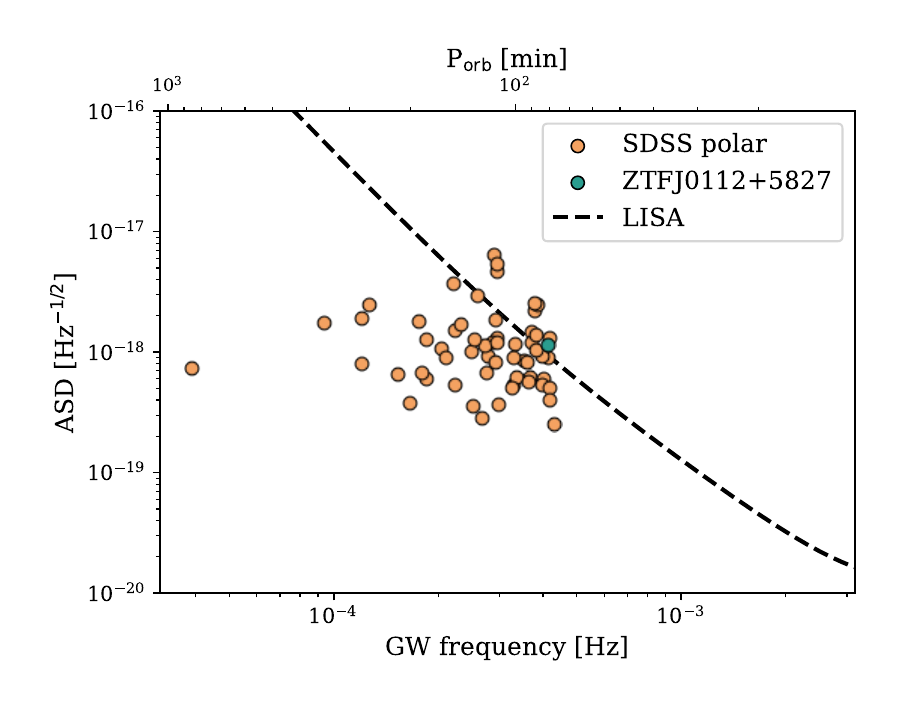}
    \caption{Amplitude spectral density (ASD) calculated over four years of LISA operation, as a function of GW frequency, is shown. $T_{\text {obs }}$ is the observation duration and $\mathcal{A} \sqrt{T_{\text {obs }}}$ is the ASD. The corresponding LISA instrument noise (excluding the galactic foreground) is depicted by the dashed black line. The SDSS polar sample from \citet{inightCatalogueCataclysmicVariables2023} is represented by orange circles, with ZTF J0112+5827 highlighted by a blue circle.}
    \label{fig:LISA}
\end{figure}

In this section, we evaluate whether ZTF J0112+5827 could be a potential detectable GW source for future space-based GW detection missions. The orbital period of ZTF J0112+5827 is 80.9 minutes, which corresponds to a GW frequency of 0.4 mHz \citep{2023scaringi}, situated within the detectable range of LISA. The detection of binary systems by LISA depends on the GW amplitudes of each source, which are averaged over the LISA constellation orbit and depend on the system's orbital inclination, the masses of the binary components, the orbital period, and the distance to the source. The GW frequency $f$ for CVs in general is related to the orbital period $P_{\text{orb}}$ by the equation

\begin{equation} f = \frac{2}{P_{\text{orb}}}. \end{equation}

The corresponding GW amplitude $\mathcal{A}$ is expressed as

\begin{equation} \mathcal{A} = \frac{2(G\mathcal{M})^{5/3}}{c^4d}(\pi f)^{2/3}, \end{equation}
where $G$ denotes the gravitational constant, $c$ represents the speed of light, $d$ is the distance to the binary system, and $\mathcal{M}$ is the chirp mass. The chirp mass is defined in terms of the binary component masses $m_1$ and $m_2$ as follows:
\begin{equation} 
\mathcal{M} = \frac{(m_1m_2)^{3/5}}{(m_1 + m_2)^{1/5}}.
\end{equation}

As we have not observed any eclipse in ZTF J0112+5827, we cannot determine the masses of its components; hence, we can only estimate their masses based on other observations. The WD mass in magnetic CVs is generally higher than in nonmagnetic CVs \citep{2015ferrario}. Studies have utilized eclipses to determine WD masses, observing that in polars with periods below 120 minutes, the WD mass is typically above 0.8 solar masses \citep[e.g.,][]{2019ecl_model,2016kotze,2023kochkina,rodriguezDiscoveryTwoPolars2023}. Therefore, we assume that the WD mass of the ZTF J0112+5827 is $m_1 = 0.80\,M_\odot$. We conclude that ZTF J0112+5827 is not a period bouncer. Therefore, we assign the corresponding donor star's mass, $m_2$, based on the revised evolutionary track from \citet{knigge2011}, which yields 0.07 $M_\odot$. ZTF J0112+5827 has a period of 80.912 min and a distance of $\sim$364 pc. Our evaluation is that it is possibly a potential gravitational source that is detectable by LISA. In Fig. \ref{fig:LISA}, we show the expected ASD of the ZTF J0112+5827 calculated over four years of LISA operation, as well as the corresponding LISA detection limit. As a comparison, we have also shown the ASD for polars observed through SDSS from \citet{inightCatalogueCataclysmicVariables2023}. We employed the same mass-estimation method for these samples as we did for ZTF J0112+5827. However, this estimation is entirely based on general assumptions regarding $m_1$ and $m_2$. Future determinations of the component masses of ZTF J0112+5827 would significantly aid in examining its potential to be detected by LISA for GW signals.

\section{Conclusion}

In this paper, we report a newly identified polar system, ZTF J0112+5827.  We determined its orbital period to be 80.9 minutes by analyzing its light curve. We demonstrate that the peculiar double spikes of its light curve can be explained by the cyclotron radiation emitted by charged particles on the WD surface. We obtained its phase-resolved spectra, and by fitting the characteristics of its cyclotron harmonic we determined that the surface magnetic-field strength of the WD is approximately $B \approx$  38.7 MG. 

We derived the system's internal accretion structure by analyzing the Doppler tomograms of its emissions. As expected, there is no accretion disk within the binary system of ZTF J0112+5827. The main emissions come from the accretion stream and the magnetic-field lines near the surface of the WD, which further confirms that it is a polar system. Spectral analysis reveals significant Balmer and helium-emission lines, showing phase-dependent variations consistent with changes in the visibility of the accretion region.

Albeit speculatively, we hypothesize that if the WD in ZTF J0112+5827 has a mass comparable to the average mass of known polars, it has likely not yet transitioned through the periodic bouncing phase. Consequently, its more massive companion may enable it to become a GW source detectable by LISA in the future.

\section*{Data availability} 
The supplementary appendices are available at \href{https://doi.org/10.5281/zenodo.14591501}{https://doi.org/10.5281/zenodo.14591501}.

\begin{acknowledgements}
Acknowledgements. This project is supported by the National Key R\&D Program of China (2020YFC2201400). J.L. and C.L. are supported by the National Natural Science Foundation of China (NSFC, grant No. 12233013). X.W. also acknowledges receiving the Tencent Xplorer Prize. N.E.R. acknowledges support from PRIN-INAF 2022 ("Shedding light on the nature of gap transients: from the observations to the models"). T.C. acknowledges funding from the support of the National Natural Science Foundation of China (NSFC, grant No. 12273002). H.G. acknowledges the support of the National Natural Science Foundation of China (NSFC, grant Nos. 12288102, 12173081), the National Key R\&D Program of China (grant No. 2021YFA1600403), and the Key Research Program of Frontier Sciences of the Chinese Academy of Sciences (CAS, grant No. ZDBS-LY-7005). W.J.H. and P.H.T. acknowledge the support of the NSFC (grant No. 12273122). We acknowledge the science research grants from the China Manned Space Project.This work is based on observations made with the Gran Telescopio Canarias (GTC), installed at the Spanish Observatorio del Roque de los Muchachos of the Instituto de Astrofísica de Canarias on the island of La Palma.
\end{acknowledgements}

\bibliographystyle{aa}

\bibliographystyle{aasjournal}

\begin{thebibliography}{99}
\expandafter\ifx\csname natexlab\endcsname\relax\def\natexlab#1{#1}\fi

\bibitem[Abril et al.(2020)]{2020Abril}
Abril, J., Schmidtobreick, L., Ederoclite, A., et al. 2020, MNRAS, 492, L40

\bibitem[{{Amaro-Seoane} et al.(2023)}]{2023amaro} Amaro-Seoane, P., Andrews, J., Arca Sedda, M., et al.\ 2023, Living Reviews in Relativity, 26, 2

\bibitem[{{Amaro-Seoane} et~al.(2017)}]{2017amaro}
{Amaro-Seoane}, P., {Audley}, H., {Babak}, S., et al. 2017, arXiv:1702.00786


\bibitem[{{Ag{\"u}eros} et~al.(2009)}]{2009agueros}
{Ag{\"u}eros}, M.~A., {Anderson}, S.~F., {Covey}, K.~R., et~al. 2009, \apjs, 181, 444

\bibitem[{{Balman} et~al.(2024)}]{2024balman}
{Balman}, {\c S}., {Khamitov}, I., {Kolbin}, A., et~al. 2024, \aap, 684, A190

\bibitem[{{Bellm} et~al.(2019)}]{bellm2019}
{Bellm}, E.~C., {Kulkarni}, S.~R., {Graham}, M.~J., et~al. 2019, \pasp, 131, 018002.

\bibitem[{{Bera} \& {Bhattacharya}(2017)}]{2017bera}
{Bera}, P., \& {Bhattacharya}, D. 2017, \mnras, 465, 4026

\bibitem[{{Boller} et~al.(2016)}]{2016rosat3}
{Boller}, T., {Freyberg}, M.~J., {Tr{\"u}mper}, J., et~al. 2016, \aap, 588, A103

\bibitem[{{Borisov} et~al.(2015)}]{2015borisov}
{Borisov}, N.~V., {Gabdeev}, M.~M., {Shimansky}, V.~V., et~al. 2015, Astron. Lett., 41, 646

\bibitem[{{Breytenbach} et~al.(2019)}]{2019ecl_model}
{Breytenbach}, H., {Buckley}, D.~A.~H., {Hakala}, P., et~al. 2019, \mnras, 484, 3831

\bibitem[{{Campbell} et~al.(2008)}]{2008campbella}
{Campbell}, R. K., {Harrison}, T. E., {Schwope}, A. D., \& {Howell}, S. B. 2008, Astrophys. J., 672, 531

\bibitem[{Chanmugam}(1980)]{1980chan} {Chanmugam}, G. 1980, \apj, 241, 1122

\bibitem[{{Cropper}(1990)}]{1990cropper}
{Cropper}, M. 1990, \ssr, 54, 195

\bibitem[{{Das} \& {Mukhopadhyay}(2013)}]{2013das}
{Das}, U., \& {Mukhopadhyay}, B. 2013, \prl, 110, 071102

\bibitem[{{D{\'e}k{\'a}ny} et~al.(2020)}]{dekanyztf}
{D{\'e}k{\'a}ny}, R., {Smith}, R.~M., {Riddle}, R., et~al. 2020, \pasp, 132, 038001.

\bibitem[{{Faulkner}(1971)}]{1971faulkner}
{Faulkner}, J. 1971, \apj, 170, L99

\bibitem[{{Ferrario} et~al.(1992)}]{1992ferrario}
{Ferrario}, L., {Wickramasinghe}, D.~T., {Bailey}, J., {Hough}, J.~H., \& {Tuohy}, I.~R. 1992, \mnras, 256, 252--260

\bibitem[{{Ferrario} et~al.(2015)}]{2015ferrario}
{Ferrario}, L., {de Martino}, D., \& {G{\"a}nsicke}, B.~T. 2015, Space Sci. Rev., 191, 111

\bibitem[{{Fuchs} et~al.(2016)}]{2016fuchs}
{Fuchs}, J.~T., {Dunlap}, B.~H., {Dennihy}, E., et~al. 2016, \mnras, 462, 2382

\bibitem[{{Galiullin} et~al.(2024)}]{2024galiullin}
{Galiullin}, I., {Rodriguez}, A.~C., {Kulkarni}, S.~R., et~al. 2024, \mnras, 528, 676

\bibitem[{{G{\"a}nsicke} et~al.(2001)}]{gansicke2001}
{G{\"a}nsicke}, B.~T., {Fischer}, A., {Silvotti}, R., \& {de Martino}, D. 2001, \aap, 372, 557

\bibitem[{{G{\"a}nsicke} et~al.(2009)}]{2009gansicke}
{G{\"a}nsicke}, B.~T., {Dillon}, M., {Southworth}, J., et~al. 2009, \mnras, 397, 2170

\bibitem[{{Graham} et~al.(2019)}]{graham2019}
{Graham}, M.~J., {Kulkarni}, S.~R., {Bellm}, E.~C., et~al. 2019, \pasp, 131, 078001

\bibitem[{{Hameury} \& {Lasota}(2017)}]{2017hameury_dn_ip}
{Hameury}, J.-M., \& {Lasota}, J.-P. 2017, \aap, 602, A102

\bibitem[{{Harrop-Allin} et~al.(1999)}]{1999huaquarii}
{Harrop-Allin}, M.~K., {Cropper}, M., {Hakala}, P.~J., et~al. 1999, \mnras, 308, 807

\bibitem[{{Hellier}(2001)}]{hellierbook}
{Hellier}, C. 2001, Cataclysmic Variable Stars (Chichester: Praxis Publishing)

\bibitem[{{Howell} et~al.(2006)}]{2006andrewhowell}
{Howell}, D.~A., {Sullivan}, M., {Nugent}, P.~E., et~al. 2006, \nat, 443 308

\bibitem[{{Huang} et~al.(2020)}]{2020PhRvD.102f3021H}
{Huang}, S.-J., {Hu}, Y.-M., {Korol}, V., et al. 2020, Phys. Rev. D, 102, 063021

\bibitem[{{Inight} et~al.(2023)}]{inightCatalogueCataclysmicVariables2023}
{Inight}, K., {G{\"a}nsicke}, B.~T., {Breedt}, E., et~al. 2023, \mnras, 524, 4867

\bibitem[{{Knigge} et~al.(2011)}]{knigge2011}
{Knigge}, C., {Baraffe}, I., \& {Patterson}, J. 2011, \apjs, 194, 28

\bibitem[{{Kochkina} et~al.(2023)}]{2023kochkina}
{Kochkina}, V. Y., {Kolbin}, A. I., {Borisov}, N. V., et al. 2023, Astron. Lett., 49, 706


\bibitem[{{Kolb} \& {Baraffe}(1999)}]{1999kolb}
{Kolb}, U. \& {Baraffe}, I. 1999, \mnras, 309, 1034

\bibitem[{{Kolbin} \& {Borisov}(2020)}]{2020kolbin}
{Kolbin}, A. I. \& {Borisov}, N. V. 2020, Astron. Lett., 46, 812


\bibitem[{{Kolbin} et~al.(2022)}]{2022kolbin}
{Kolbin}, A.~I., {Borisov}, N.~V., {Serebriakova}, N.~A., et~al. 2022, \mnras, 511, 20

\bibitem[{{Kotze} et~al.(2015)}]{2015kotze}
{Kotze}, E.~J., {Potter}, S.~B., \& {McBride}, V.~A. 2015, \aap, 579, A77

\bibitem[{{Kotze} et~al.(2016)}]{2016kotze}
{Kotze}, E.~J., {Potter}, S.~B., \& {McBride}, V.~A. 2016, \aap, 595, A47

\bibitem[{{Lindegren} et~al.(2021)}]{2021lindegren}
{Lindegren}, L., {Klioner}, S.~A., {Hern{\'a}ndez}, J., et~al. 2021, \aap, 649, A2

\bibitem[{{Littlefield} et~al.(2017)}]{littlefield2017}
{Littlefield}, C., {Garnavich}, P., {Hoyt}, T., \& {Kennedy}, M. 2017, \aj, 155, 18

\bibitem[{{Lomb}(1976)}]{Lomb1976}
{Lomb}, N.~R. 1976, \apss, 39, 447

\bibitem[{{Mason} et~al.(2019)}]{2019mason}
{Mason}, P.~A., {Wells}, N.~K., {Motsoaledi}, M., {Szkody}, P., \& {Gonzalez}, E. 2019, \mnras, 488, 2881

\bibitem[{{Masci} et~al.(2019)}]{masci_ztf}
{Masci}, F.~J., {Laher}, R.~R., {Rusholme}, B., et~al. 2019, \pasp, 131, 018003.

\bibitem[{{Mukai}(2017)}]{mukai2017}
{Mukai}, K. 2017, \pasp, 129, 062001

\bibitem[{{Oke} \& {Gunn}(1982)}]{1982oke}
{Oke}, J.~B. \& {Gunn}, J.~E. 1982, \pasp, 94, 586

\bibitem[{{Paczynski}(1981)}]{1981paczynski}
{Paczynski}, B. 1981, Acta Astronomica, 31, 1

\bibitem[{{Pala} et~al.(2020)}]{pala2020}
{Pala}, A.~F., {G{\"a}nsicke}, B.~T., {Breedt}, E., et~al. 2020, \mnras, 494, 3799-3827

\bibitem[{{Press} et~al.(2007)}]{2007nr}
{Press}, W. H., {Teukolsky}, S. A., {Vetterling}, W. T., et al. 2007, Numerical Recipes: The Art of Scientific Computing, 3rd edn. (Cambridge: Cambridge Univ. Press)


\bibitem[{{Pringle}(1981)}]{1981pringle}
{Pringle}, J.~E. 1981, \araa, 19, 137

\bibitem[{{Rappaport} et~al.(1983)}]{rappaport1983}
{Rappaport}, S., {Joss}, P.~C., \& {Verbunt}, F. 1983, \apj, 275, 713

\bibitem[Ramaty(1969)]{1969Ramaty} Ramaty, R., 1969, \apj, 158, 753

\bibitem[{{Ren} et~al.(2023)}]{2023ren}
{Ren}, L., {Li}, C., {Ma}, B., et~al. 2023, \apjs, 264, 39

\bibitem[{{Ritter} \& {Kolb}(2003)}]{2003ritter}
{Ritter}, H. \& {Kolb}, U. 2003, \aap, 404, 301

\bibitem[{{Rodriguez} et~al.(2023)}]{rodriguezDiscoveryTwoPolars2023}
{Rodriguez}, A.~C., {Kulkarni}, S.~R., {Prince}, T.~A., et~al. 2023, \apj, 945, 141

\bibitem[{{Sarkar} et al.(2023)}]{2023sarkar} Sarkar, A., Ge, H., \& Tout, C.~A.\ 2023, \mnras, 520, 3187 

\bibitem[{{Scaringi} et~al.(2023)}]{2023scaringi}
{Scaringi}, S., {Breivik}, K., {Littenberg}, T.~B., et~al. 2023, \mnras~Lett., 525, L50

\bibitem[{{Scargle}(1982)}]{Scargle1982}
{Scargle}, J.~D. 1982, \apj, 263, 835

\bibitem[{{Schmidt} et~al.(2005)}]{2005schmidt}
{Schmidt}, G.~D., {Szkody}, P., {Homer}, L., et~al. 2005, \apj, 620, 422

\bibitem[{{Schreiber} et~al.(2021)}]{schreiber2021}
{Schreiber}, M.~R., {Belloni}, D., {G{\"a}nsicke}, B.~T., {Parsons}, S.~G., \& {Zorotovic}, M. 2021, Nat. Astron., 5, 648-654

\bibitem[{{Schreiber} et~al.(2024)}]{2024schreiber}
{Schreiber}, M.~R., {Belloni}, D., \& {Schwope}, A.~D. 2024, \aap, 682, L7
\bibitem[{{Schwope}(1990)}]{1990schwope}
{Schwope}, A. D. 1990, Rev. Mod. Astron., 3, 44

\bibitem[Schwope et al.(2024)]{2024schwope}  
Schwope, A., Knau, K., Kurpas, J., et al. 2024, \aap, 690, A243  

\bibitem[{{Szkody} et~al.(2021)}]{2021szkody}
{Szkody}, P., {Olde Loohuis}, C., {Koplitz}, B., et~al. 2021, \aj, 162, 94

\bibitem[{{Takata} et~al.(2022)}]{2022takata}
{Takata}, J., {Wang}, X.~F., {Kong}, A.~K.~H., et~al. 2022, \apj, 936, 134

\bibitem[{{Tr\"umpe}(1982)}]{1986trumper}
{Tr\"umper}, J. 1982, Adv. Space Res., 2, 241-249

\bibitem[{{Verbunt} \& {Zwaan}(1981)}]{1981verbunt}
{Verbunt}, F. \& {Zwaan}, C. 1981, \aap, 100, L7

\bibitem[{{Warner}(1995)}]{warner95}
{Warner}, B. 1995, Cambridge Astrophysics Series, 28

\bibitem[{{Wickramasinghe} \& {Ferrario}(2000)}]{2000wickramasinghe}
{Wickramasinghe}, D.~T., \& {Ferrario}, L. 2000, \pasp, 112, 873

\bibitem[{{Worraker}(2001)}]{2001Worraker}
{Worraker}, B. 2001, The Astronomer, 38, 47

\end{thebibliography}

\end{document}